# Magnetic anisotropy and de Haas – van Alphen oscillations in a Bi microwire array studied via cantilever magnetometry at low temperatures

M. J. Graf,[a] C. P. Opeil,[a] and T. E. Huber [b]

[a] Department of Physics, Boston College, Chestnut Hill, MA 02467 USA

[b] Laser Laboratory, Howard University, Washington, DC 20059 USA

Abstract

We report measurements of the low temperature (T = 0.5 K) oscillatory magnetization in a high-density array of 50 μm diameter wires of polycrystalline Bi utilizing a high sensitivity silicon cantilever magnetometer. We find that the magnetic response is strongly anisotropic, being much larger for magnetic field perpendicular than for fields parallel to the wire-axis. We argue that this is a geometric effect caused by the large aspect ratio of the individual microwires in the array. The magnetic response of the microwires is dominated by the light electrons due to the larger cyclotron orbits in comparison with the heavier holes. We find that de Haas - van Alphen oscillations are easily resolved, and discuss the application of this technique to the study of Bi nanowire arrays.

## I. Introduction

Bi is a semimetal with exceptional electronic transport properties. The electron mean-free-path in single crystal Bi can be as long as a millimeter at 4.2 K, several orders of magnitude larger than for most metals.[1] The Fermi wavelength $\lambda_F$ is about 600 Å,[2] as opposed to a few Å in most metals, and as a result, boundary scattering is nearly specular.[3] These long characteristic lengths derive from its ellipsoidal Fermi surfaces, low carrier densities and small carrier effective masses.[4] Recently, the electronic transport properties of Bi nanowires of diameter $d_W$ have been investigated with the goal of understanding the physics of quantum wires ($d_W < \lambda_F$) and the semimetal-to-semiconductor transition of Bi as the wire diameter decreases below the Fermi wavelength.[5] Recently advances have been made in the fabrication of very high quality Bi wires. The Ulitovskii-Taylor method yields glass-encapsulated single-crystal individual wires in the diameter range 70 nm-20 microns.[6] Arrays and networks of wires with diameters in the range from 6 nm to several microns can be formed by pressure-injecting,[7] vacuum infiltration[8] and electrochemical deposition[9] into non-conducting host template materials. These are of particular relevance to thermoelectric applications. Bi and Bi-Sb crystals have the largest thermoelectric figure-of-merit of any material for T<100 K,[10] and it has been predicted that Bi quantum wires can be engineered to have enhanced thermoelectric properties because of the increase of the density of states resulting from quantum confinement along the wire axis.[11] Moreover, in the case of high-density arrays of nanowires one has all the favorable traits of engineered electronic properties due to confinement and doping,[12] while retaining the bulk attributes necessary for thermoelectric cooling as well as other large-scale applications. Recent measurements of the thermopower of arrays of 9 to 15-nm Bi wires by Heremans *et al.* seem to confirm that confinement enhances the thermoelectric performance[13] and this topic is currently the subject of intense investigation.[14] Additionally, sub-micron Bi wires exhibit Aharonov-Bohm (AB) oscillations, that is oscillations of the resistance as a function of magnetic field with a period $\Delta B \sim \Phi_o / d_W^2$, where $\phi_0$ is the quantum fluxoid, caused by quantum interference of electrons in "whispering gallery" modes along the wire periphery.[15] AB oscillations are potentially useful for realizing certain mathematical algorithms necessary for quantum computers.[16]

In order to successfully engineer the properties of the materials for applications, it is of utmost importance to understand the effects of confinement and doping on the Fermi surface. This is especially true for nanowire systems with a large surface-to-volume ratio, since surface effects may introduce additional carriers into the system, thus significantly altering the properties of the nanowires.[17-19] The Fermi surface is effectively probed by de Haas-van Alphen (dHvA) and Shubnikov-de Haas (SdH) oscillations.[20]

dHvA and SdH oscillations are caused by the Landau level quantization of closed orbits of charge carriers in an applied magnetic field. As the field is increased, the energy of the Landau level increases, and when its minimum value becomes equal to the Fermi energy, it is suddenly depopulated. This oscillatory behavior in the density of states manifests itself directly in the dHvA oscillations of the magnetization, and in other bulk thermodynamic properties of the material. This oscillatory behavior, which is periodic in 1/B, can also be observed in transport properties, *e.g.*, the magnetoresistance. The relaxation time for electron scattering is temporarily increased at the magnetic field value when the Landau level is depopulated, giving rise to a dip in the magnetoresistance, resulting in SdH oscillations. Characterization of the electronic properties of nanowire arrays by SdH measurements is problematic. Since SdH measurements depend on electron collisions, they can be complicated by other effects e.g., Stark quantum interference.[21] Also, one must make electrical contact to a large number of parallel Bi wires, resulting in further complications due to the properties of the contact materials, and diffraction effects at the contact/nanowire interface.[19] In contrast, dHvA measurements do not require electrical contacts, and measurements directly probe the oscillatory density of states. Theoretical analysis is fairly straightforward. Experimentally, dHvA is a superior probe of the density of states of nanowire arrays when compared to SdH. However, we are not aware of any studies of the magnetic properties of confined Bi. As we will show such studies are challenging due to the small Bi sample mass, complications introduced by demagnetization factors, and wire size effects. High-precision magnetometry is required to extract the oscillatory behavior.

In this work we report measurements of the oscillatory magnetization in a high-density array of 50 μm diameter Bi wires utilizing a high sensitivity silicon cantilever magnetometer. The very large aspect ratio of the individual microwires results in a strongly anisotropic magnetic response. We find that the magnetic response of the microwires is dominated by the light electrons due to the larger cyclotron orbits in comparison with the heavier holes. dHvA oscillations are easily resolved, and show that the technique can readily be extended to measure the Fermi surface properties of nanowire arrays.

## II. Experimental

To test the viability of cantilever magnetometry for nanowire arrays we have utilized an array of 50 µm diameter wires in a borosilicate glass host template (Galileo, Sturbridge, MA). The template is not magnetic. The Bi wire is synthesized by high-pressure injection of the melt into the template.[22] The large wire diameter was chosen to minimize the potential complications due to confinement effects[23] while maintaining the high degree of geometrical anisotropy present in nanowire arrays, and because one can make electrical contact to the large wires relatively easily using silver epoxy, thus enabling a comparison between dHvA and SdH measurements. X-ray diffraction on this sample revealed that the Bi microwires grown in this template (with relatively large channel diameters) are polycrystalline, in contrast to the highly oriented nanowires fabricated by us for our previous studies.[18,19,22] The sample was cut into a parallelopiped with approximate dimensions 0.4mm x 1mm x 1.7 mm with a total mass of 5.0 mg, and is shown as viewed along the microwire axes in Fig. 1. The array is oriented such that the individual microwire lengths are 1.7 mm. There are about 120 microwires in the sample studied, and the estimated total volume and mass of Bi are $4.1 \times 10^{-10}$ m$^3$ and 4.0 mg, respectively. Individual microwires have a large aspect ratio of approximately 34. Assuming that the magnetization of the glass host is negligible, and neglecting the inter-wire magnetic correlations,[24] the large microwire aspect ratio results in demagnetization factors of 0.002 and 0.499 for magnetic fields applied parallel and perpendicular to the wire axis, respectively.[25]

Cantilever magnetometers have been developed to provide high precision measurements in strong magnetic fields, and the technique has been described elsewhere in detail.[26] Here we will summarize the basic operating principles. As shown in Fig. 2, a sample is placed on a flexible cantilever, and the deflection of the cantilever can be monitored very precisely by measuring the capacitance between the cantilever and a fixed reference plate. The cantilever is placed along the central axis of a superconducting solenoid at a point away from the field center, and the cantilever's surface normal is parallel to this central axis. The sample is thus located in an applied magnetic field **H** with a non-zero gradient along the axis, and acquires a magnetization **M** in the field. The resulting deflection of the cantilever contains contributions from the magnetic force $\mathbf{F_c} = \nabla(\mathbf{M} \cdot \mathbf{H})$ and, if the sample is magnetically anisotropic (i.e., **M** not parallel to **H**), a torque $\boldsymbol{\tau_c} = \mathbf{M} \times \mathbf{H}$. If **M** is linearly proportional to the magnitude of **H**, both the torque and force terms vary as H$^2$. When the cantilever deflection is small the fractional change in capacitance relative to the zero magnetic field value, $\delta C \equiv \Delta C(B)/C(0)$, should also vary as $H^2$. In the absence of a torque contribution, $\delta C$ is proportional to the magnetic susceptibility.

In this work we have utilized a micro-machined Si cantilever mounted in a $^3$He refrigerator. This particular cantilever has been used to probe magnetic transitions in bulk strongly correlated electron systems.[27] The magnetic force response of the cantilever was calibrated by putting a fixed current through a circuit loop of known area deposited on the cantilever, and measuring the subsequent deflection in an applied magnetic field and a known field gradient. As shown in Fig. 2, the experimental geometry is such that a positive value of $\delta C$ is characteristic of diamagnetic response in $F_c$: the sample is pushed away from field center (down), thereby increasing the capacitance. The 50 μm wire array was fixed to the cantilever by Apiezon N-grease. In the following discussion of the results, "longitudinal" denotes that the sample was mounted with the microwire axes approximately aligned with the magnetic field, while "transverse" means that the same sample was rotated from this configuration by 90º so that the field was aligned perpendicular to the axes. The cantilever apparatus itself is fixed for all experiments. Although the samples are polycrystalline, the high aspect ratio will introduce magnetic anisotropy into the system. If the microwires are not aligned exactly parallel or perpendicular to the applied magnetic field, this anisotropy will produce a magnetization along microwire axis and a torque will result: the microwires will tend to align along the applied magnetic field.

**III. Results**

In Figure 3 the low-field response for the cantilever is shown for both the longitudinal and transverse modes at a temperature of 0.5 K. Both are seen to vary as approximately $H^2$, as expected. Bismuth has a low-temperature, low-field anisotropic susceptibility (per volume) that ranges between $-1.2 \times 10^{-5}$ and $-1.8 \times 10^{-5}$.[28] The capacitance change is roughly 0.043 femtofarads in magnitude in an applied magnetic field of 0.3 T, as shown in Fig. 3 (the zero magnetic field capacitance for our device is 0.43 pF). Assuming the response is solely due to the magnetic force (i.e., no torque), our cantilever calibration factor yields a susceptibility with magnitude of $5 \times 10^{-6}$. This is in order-of-magnitude agreement with the accepted value; we believe the discrepancy arises from a significant torque contribution to the magnetic signal, as discussed below.

Because Bi is diamagnetic, $\delta C$ should be positive. Instead, we see that the signs for $\delta C$ in the longitudinal and transverse modes are different, indicating that there is a contribution to the overall deflection from the torque term $\tau_c$, which is comparable to that from the force term. This is believed to result from the fact that the magnetic field is not oriented exactly parallel or perpendicular to the microwire wire axes in the longitudinal and transverse modes, respectively. The misalignment results in a torque that tends to align the microwires along the magnetic field direction. The conjecture is supported by the high-field results shown in Fig. 4. The longitudinal response has changed from negative to positive, although the large amplitude of the dHvA oscillations makes an unambiguous determination of a crossover difficult. This behavior is consistent with an alignment of the microwires for magnetic fields above roughly 3 Tesla: the torque term has become small, and the response is dominated by the diamagnetic force term. This is to be contrasted with the transverse response, which shows a dramatic increase at high fields. In this case a sample reorientation of roughly $90^o$ is required to align the microwires with the field. Because the torque term does not saturate at intermediate fields it can become quite large at high fields. This is clearly observed, with an almost 60% change in the capacitance – far out of the linear mechanical response regime of the cantilever – as compared to a change of about 1% for the high field longitudinal response.

The dHvA oscillations are readily observed, and we highlight these in Fig. 5 by plotting the magnetic field derivative of $\delta C$ taken at $T = 0.5$ K versus inverse magnetic field over the range where the oscillations are observable. The data have been slightly offset for clarity, and the dramatic increase in the transverse data at very high fields has been left off the plot. The dominant oscillations appear to be out of phase, which we believe is due to the significant torque contribution to the signal, consistent with the opposite signs observed for the low-field data in Fig. 3. Data was also taken at 4.2 K in both transverse and longitudinal configurations, and no appreciable damping of the oscillation amplitude was observed.

The resulting dHvA spectra are complicated, and in addition the major maxima and minima there are also several inflection points. Since the microwires are polycrystalline it would be extremely difficult to identify all the oscillatory periods; moreover such an effort would be beyond the scope of the work presented here. Nonetheless, we have assigned indices to the major extrema (maxima being assigned integer values, and minima half-integer), and find that both orientations yield a primary dHvA period of 0.42(1) $T^{-1}$. This data is shown in Figure 6.

We also have measured the longitudinal and transverse magnetoresistance for the sample studied in this work. Electrical contacts were made using silver epoxy, and the measured resistance included a dominant contribution from the contact resistance of about 0.5 Ω between the silver epoxy and the microwire array; the approximate zero-field resistance of the microwires is on the order of 100 μΩ. The measurements were made using an AC resistance bridge with the samples immersed in liquid helium at 2 K. SdH oscillations were clearly discernible, and are compared to the dHvA oscillations at high fields in Fig. 7.

## IV. Discussion

First we discuss the dominant dHvA period observed for our polycrystalline microwires, which is shown in Fig. 6 to be approximately 0.42(1) T$^{-1}$. Both electron and hole pockets of the Fermi surface of Bi can contribute to dHvA. Confinement effects can be considered to have negligible effect on the Fermi surfaces and carrier effective masses for 50-micron diameter wires. For bulk Bi, the electron periods range between 0.7 T$^{-1}$ and 0.166 T$^{-1}$ for single-crystals oriented along the bisectrix and the trigonal axis, respectively; the hole periods range between 0.045 T$^{-1}$ and 0.156 T$^{-1}$ for orientations perpendicular and parallel to the trigonal axis.[29] Thus, the long period observed in this work is consistent with a contribution from electrons rather than holes. The large diamagnetic susceptibility for bulk Bi is believed to result from the orbital response (Landau-Peierls contribution) of the light electrons, averaged over the Fermi surface.[30] This picture is consistent with our observation of a predominant electron contribution to the dHvA oscillations.

The dHvA oscillation amplitude was observed to be essentially independent of temperature below 4.2 K. When the thermal energy is much lower than the cyclotron energy, $k_B T \ll \hbar \omega_C$, the amplitude of the dHvA oscillations is approximately[31]

$$A_{osc} \propto \frac{B^{1/2}}{m^*} \exp(-2\pi\Gamma/\hbar\omega_c) , \qquad (1)$$

where $B$ is the magnetic field and $\Gamma$ is the intrinsic width of the Landau levels, typically caused by impurity scattering. This width can be used to define the Dingle temperature $T_D \equiv \Gamma/\pi k_B$. At higher temperatures, the spread of the Fermi distribution causes additional damping of the dHvA amplitude, and Eq. (1) becomes

$$A_{osc} \propto \frac{T}{B^{1/2}} \exp(-2\pi^2 k_B (T+T_D)/\hbar \omega_c) . \qquad (2)$$

When $T < T_D$ the dHvA oscillation amplitude is essentially independent of temperature. Thus, for our microwire array, $T_D \geq 4.2$ K. For comparison, SdH oscillations in 10 μm thick single crystal films have a Dingle temperature of order 0.5 K.[32] If $\Gamma$ is dominated by scattering, we can write $T_D = \hbar^2 k_F / k_B l m*$, where $k_F$ is the Fermi wavevector, $l$ is the mean-free-path, and $m*$ is the effective mass. The relatively high value for $T_D$ may be explained by the dominant contribution to the magnetization of our polycrystalline microwires from the light electrons, as described in the preceding paragraph, since $T_D \propto 1/m*$. However, measurements of $T_D$ for the electronic contribution to the dHvA oscillations in bulk single crystals yields $T_D \approx 0.7$ K.[20] Because the microwires are polycrystalline, grain boundary scattering can reduce $l$, and this may also contribute to the large value of $T_D$. This speculation is tentative, however. A large value of $T_D$ would significantly damp the dHvA signal, yet our signal is quite strong, although this may in part be explained by the high sensitivity of our magnetometer.

Regarding the SdH oscillations, we note that the dominant period is fairly short (approximately 0.2 T$^{-1}$) and is most likely due to a combination of the signal due to hole and electron pockets where holes dominate. Comparison between the magnetic and the magnetoresistance signal therefore indicates that SdH is more sensitive to holes whereas dHvA is more sensitive to electrons. This implies that in those cases where both methods can be employed, they provide complementary information on the Fermi surface.

Finally, we discuss extending the cantilever technique for measuring dHvA oscillations to nanowire arrays. Magnetic measurements on nanowire arrays present additional challenges with regard to the sample size, which is typically very small, and to the expected anisotropy that is expected to be larger than that of microwire samples of polycrystalline Bi. For a conservative estimate, a typical high-density nanowire template (e.g., the 30 and 200 nm templates studied in Ref. 18) can have dimensions 3mm x 3mm x 20μm, with packing fractions on the order of 50%. Thus the total volume of Bi is 1x10$^{-10}$ m$^3$, and a total Bi mass of about 1 mg. This is approximately 25% of the microwire mass studied in this work. The resolution shown in Fig. 4 clearly can tolerate a factor of four change in the signal-to-noise ratio and still provide accurate resolution of the dHvA oscillations. Moreover, the sensitivity can be improved upon either by increasing the template thickness, or by using a thinner cantilever,[33] or both. The nanowires would have aspect ratios of 100 and 670, and so would exhibit the same high degree of magnetic anisotropy as those studied in this work.

Additional complications in measuring longitudinal dHvA oscillations for nanowire arrays need to be considered. Confinement can introduce additional anisotropy to, and a reduction in the magnitude of, the magnetic susceptibility due to the restriction of the allowed orbit sizes. The mean-free-path of the carriers, and in particular that of the light electrons in Bi, is very long. The allowed orbits of the carriers that will contribute to the diamagnetic screening must be fully contained in the wire.[30] When the magnetic field is parallel to the wire axis, the screening orbits are limited to a diameter $d_o$ smaller than the wire diameter $d_w$. For magnetic field perpendicular to the wire length orbits are less constrained, and can take the form of an ellipsoid with a minor axis smaller than the wire diameter and major axis of arbitrary length in the direction of the wire axis. This effect will be significant at low magnetic fields where the cyclotron radius is larger than the nanowire diameter. The crossover magnetic field[15] $B_c = \hbar k_F / e d_W$, where $e$ is the electron charge, is estimated to be 0.02 T for our 50 µm array (i.e., negligible), but is 1 T for 200-nm nanowires.[19] Thus for the smallest nanowires we can expect a confinement-induced reduction in the magnetic response over a significant range of magnetic field. Additionally, dHvA oscillations are not well-defined for $B < B_c$, and so we expect that nanowires with d < 200 nm may require much stronger magnetic fields than those used in this study in order to improve the dynamic range of the dHvA spectrum.

## V. Summary

In conclusion, we have demonstrated that the highly sensitive technique of cantilever magnetometry can be successfully used to study the magnetic properties of an array of 50 µm diameter Bi wires. The very large aspect ratio of the individual microwires results in a strongly anisotropic magnetic response. The magnetic response of the microwires is dominated by the light electrons due to the larger cyclotron orbits in comparison with the heavier holes. We find that the de Haas-van Alphen oscillations are easily resolved, and show that the technique can readily be applied to nanowire arrays, with the goal of understanding how the Fermi surface is modified by quantum confinement effects. We are now undertaking studies of 200 nm Bi arrays to complement our recent Shubnikov-de Haas studies.[18,19]

**Acknowledgments**

MJG and CPO would like to thank Prof. M. J. Naughton for technical advice regarding the cantilever magnetometer. The work was supported by the Division of Materials Research of the U.S. National Science Foundation under grant NSF-0072847 and by the Division of Materials of the U.S. Army Research Office under grant DAAD4006-MS-SAH.

**Figure Captions**

Fig. 1  Picture of the sample studied in this work, viewed along the microwire axes. The sample height is about 1.7 mm.

Fig. 2  Schematic diagram of the silicon cantilever magnetometer, showing the sample mounted in the (a) longitudinal and (b) transverse measurement modes. Note that the cantilever itself remains fixed in position, and the two modes are attained by remounting the same sample piece.

Fig. 3  Fractional change in capacitance versus the square of the magnetic field at weak magnetic fields in the transverse and longitudinal configurations. The temperature is 0.5 K.

Fig. 4  Fractional change in capacitance versus magnetic field at strong magnetic fields in the transverse and longitudinal configurations. The temperature is 0.5 K.

Fig. 5  Magnetic field derivative of the fractional change in capacitance versus inverse magnetic field in the transverse and longitudinal configurations. The temperature is 0.5 K. Arrows denote extrema used for a rough analysis of de Haas- van Alphen oscillations.

Fig. 6  Location in inverse field of extrema from Fig. 5 versus assigned index. Integer values correspond to maxima, while half-integers correspond to minima.

Fig. 7  Comparison of the magnetoquantum oscillations in magnetization (dHvA) and resistance (SdH) taken for the same sample with field oriented parallel (longitudinal) and perpendicular (transverse) to the microwire axes. The temperature of the dHvA curves is 0.5 K, while for the SdH curves it is 2 K. The curves have been scaled to appear of comparable size, and offset for clarity.

**Figure 1**

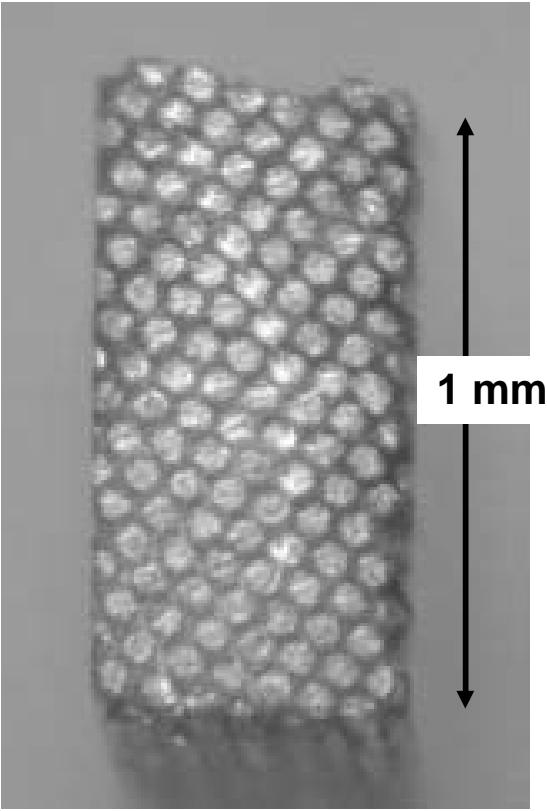

**Figure 2**

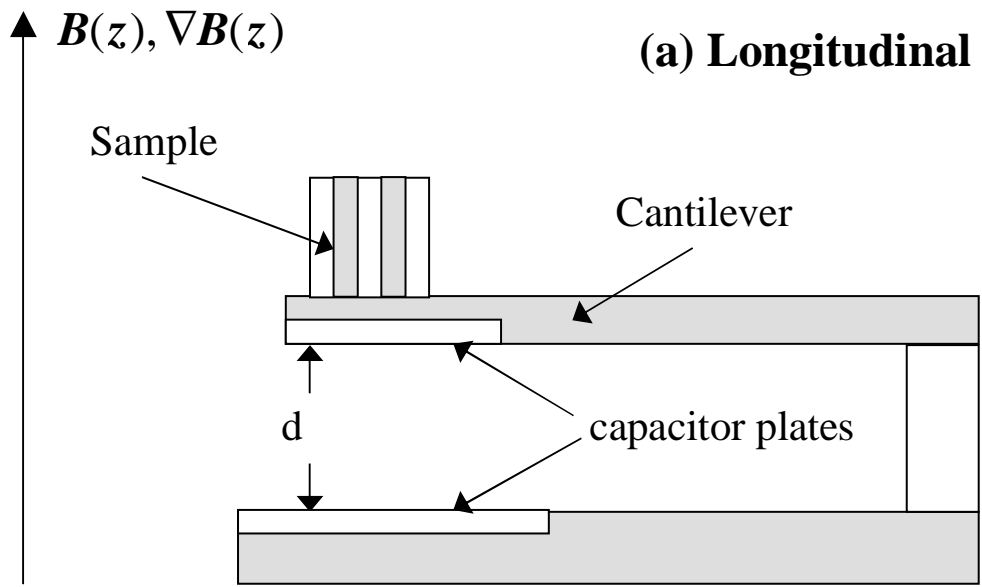

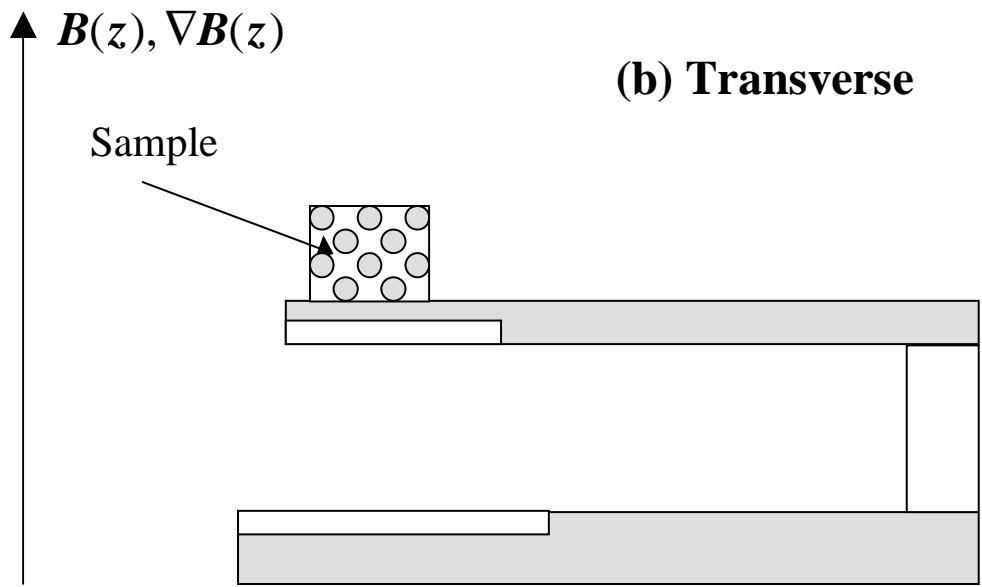

**Figure 3**

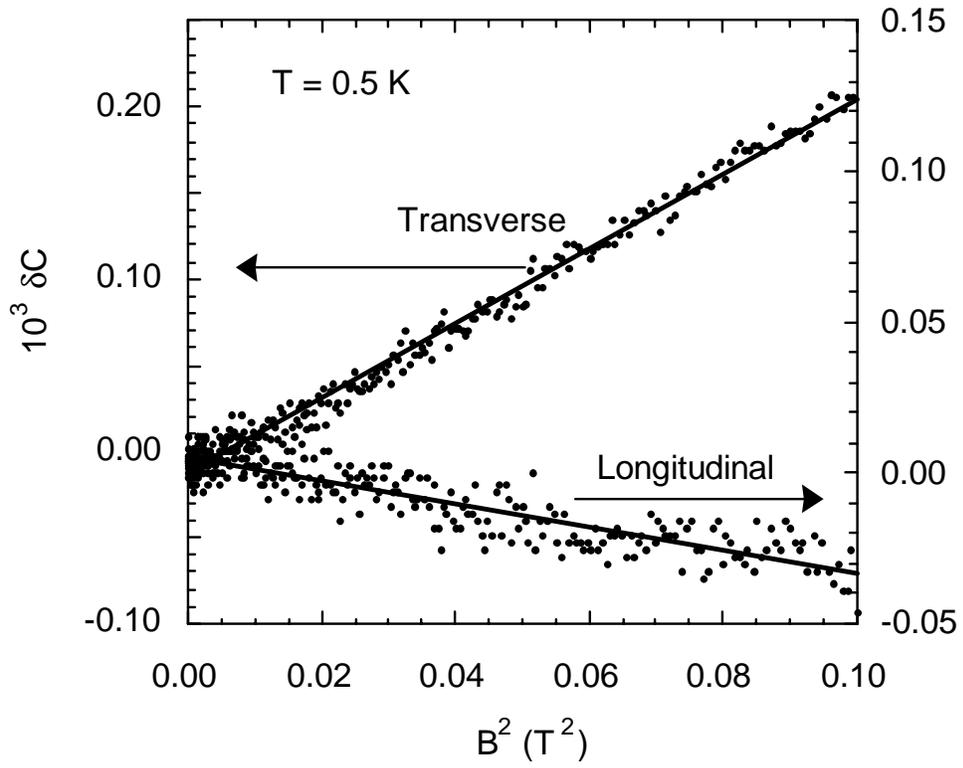

**Figure 4**

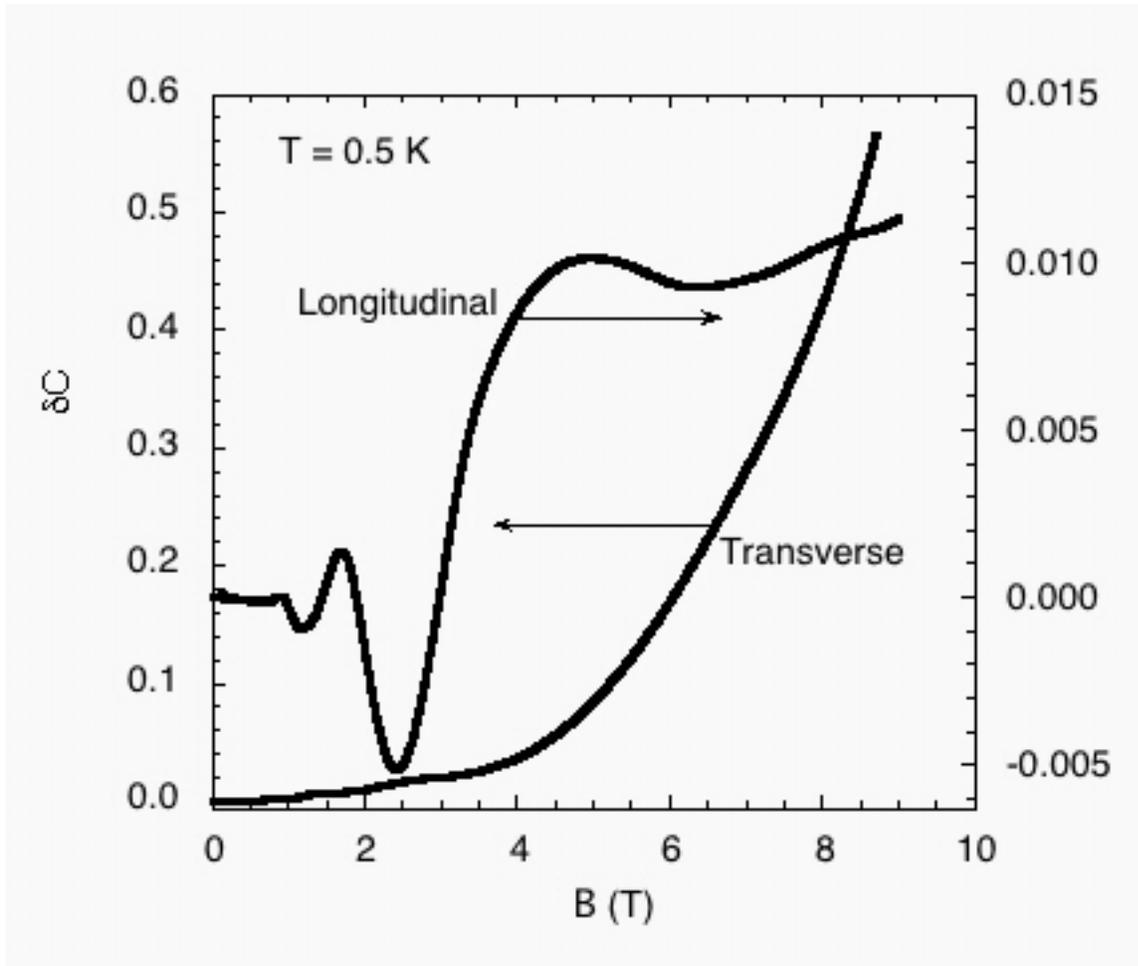

**Figure 5**

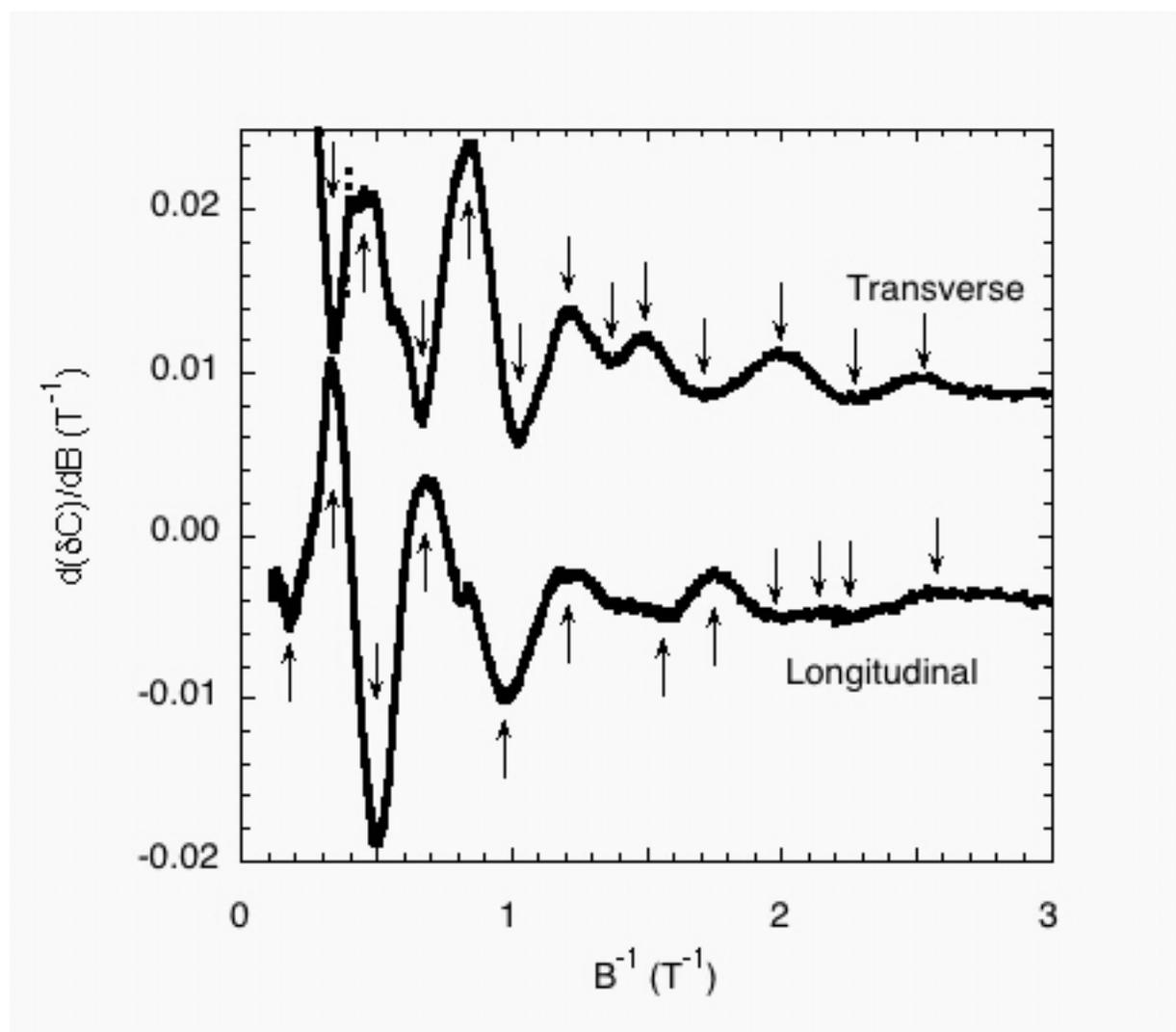

**Figure 6**

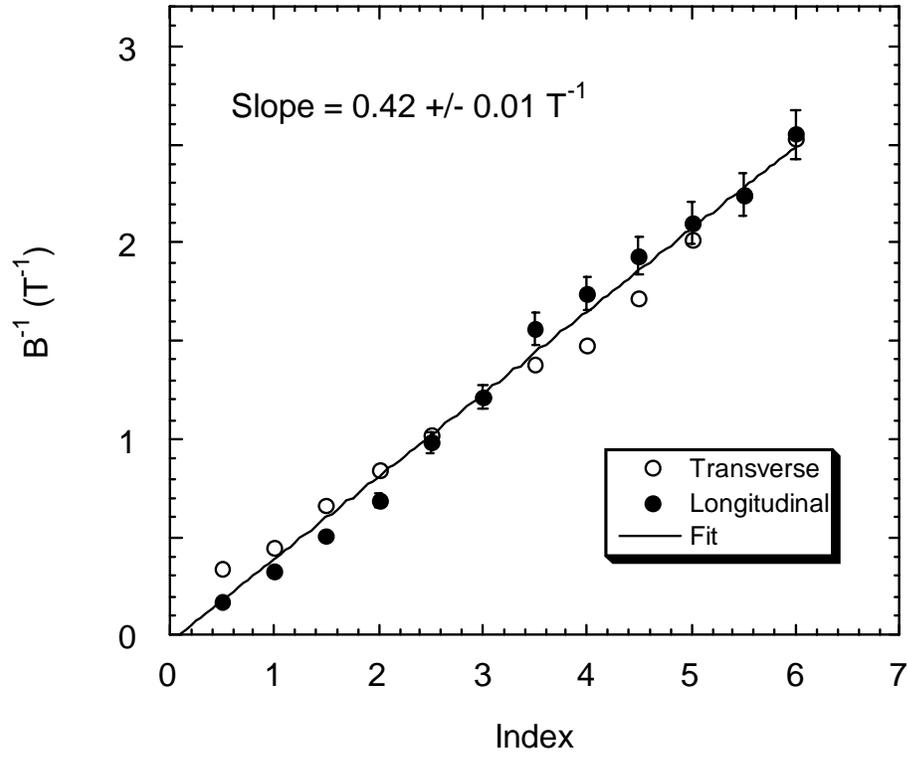

**Figure 7**

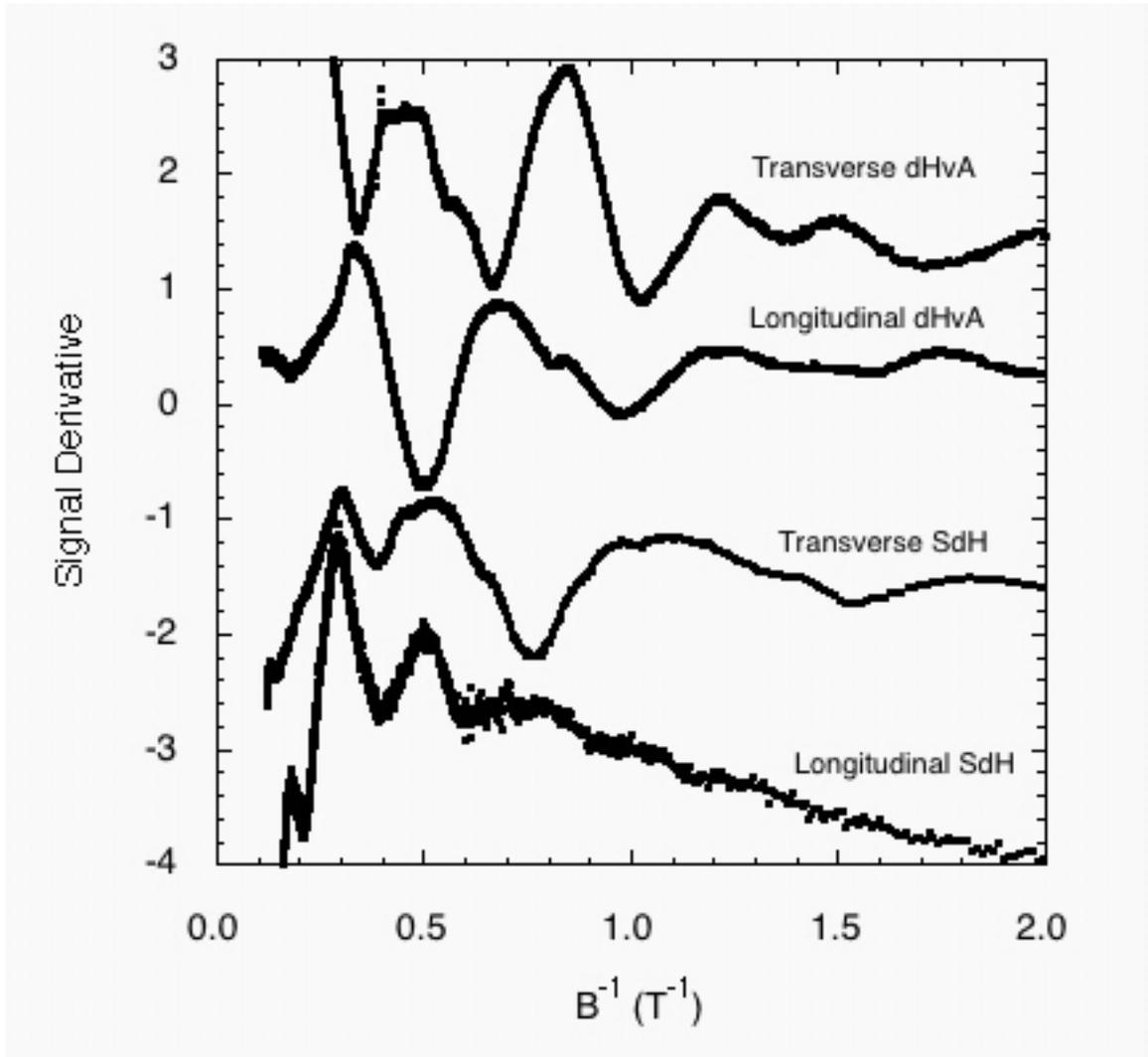